# Photon trajectories, anomalous velocities, and weak measurements: A classical interpretation

Konstantin Y. Bliokh[1,2], Aleksandr Y. Bekshaev[3],
Abraham G. Kofman[1,4], and Franco Nori[1,4]

[1]*Advanced Science Institute, RIKEN, Wako-shi, Saitama 351-0198, Japan*
[2]*A. Usikov Institute of Radiophysics and Electronics, NASU, Kharkov 61085, Ukraine*
[3]*I. I. Mechnikov National University, Dvorianska 2, Odessa, 65082, Ukraine*
[4]*Physics Department, University of Michigan, Ann Arbor, Michigan 48109-1040, USA*

Recently, Kocsis *et al.* reported the observation of "average trajectories of single photons" in a two-slit interference experiment [*Science* **332**, 1170 (2011)]. This was possible by using the quantum weak-measurements method, which implies averaging over many events, i.e., in fact, a multi-photon limit of classical linear optics. We give a classical-optics interpretation to this experiment and other related problems. It appears that weak measurements of the local momentum of photons made by Kocsis *et al.* represent measurements of the Poynting vector in an optical field. We consider both the real and imaginary parts of the local momentum, and show that their measurements have been realized in classical optics using small probe particles. We also examine the appearance of "anomalous" values of the local momentum: either negative (backflow) or exceeding the wavenumber (superluminal propagation). These features appear to be closely related to vortices and evanescent waves. Finally, we revisit a number of older works and find examples of photon-trajectories and anomalous-momentum measurements in various optical experiments.

PACS: 42.50.Xa, 42.50.Tx, 42.50.Wk, 03.65.Ta

## 1. Introduction

Two years ago Kocsis *et al*. [1] reported the experimental observation of the "averaged trajectories of single photons" in a two-slit interference experiment. This work caused enormous interest (Physics World selected it as the top breakthrough in physics in 2011) because it seemingly overcame fundamental restrictions of quantum mechanics. Indeed, simultaneous *strong* measurements of the which-path information and interference picture are impossible in standard quantum theory, like the simultaneous determination of the coordinates and momentum of the particle [2]. This is because a strong measurement of one quantity destroys the information about its complementary quantity. However, quantum *weak measurements*, introduced by Aharonov *et al.* [3–5], imply a weak coupling between the measuring and the measured sub-systems and do not perturb significantly the measured part. This approach allows the simultaneous determination of complementary quantities, albeit *averaged over many events*. Furthermore, the *weak values* of the measured quantity can be "anomalous", i.e., beyond the range of the spectrum of the corresponding Hermitian operator [3] and even complex [6]. Weak values are closely related to complex *conditional probabilities* for the system to be found in a certain state under the condition that it is subsequently found in another state [7].

Soon after the introduction of the theoretical concept of weak measurements, this method was applied to observations of tiny polarization-dependent wave shifts in classical optics [8,9]. In these experiments, two weakly-coupled degrees of freedom of light (the polarization and spatial distribution) played the roles of the measured and measuring ("meter") sub-systems. For instance, the spin-Hall effect of light (a spin-dependent shift of optical beams) [9,10] was treated as the shift of the "meter" which 'measured' the photon spin (helicity) [11]. As a result, the shift was significantly amplified owing to the anomalous weak value of the spin, which was complex and far outside its normal [−1,1] range ($\hbar = 1$ units are used throughout the paper). Despite the quantum weak-measurement formalism, experiments [8,9] clearly dealt with *classical* optical systems and all results, including the spin-Hall effect of photons, allowed fully classical descriptions [10]. This is because weak measurements require averaging over many one-particle events, so that in linear photonic systems this approach implies the classical-optics limit.

Kocsis *et al*. emphasized that their experiment [1] was realized in the quantum one-photon regime. Still, the measurements represented an average over many events, and the *same* results would appear in the continuous-wave regime [2,12]. Nonetheless, the results of [1] have not been given a classical interpretation, which would provide additional insight, complement the quantum description, and reveal interesting links. Indeed, quantum-mechanically, Kocsis *et al*. measured the *Bohmian trajectories* of photons [13]. These trajectories are determined by the *streamlines of the probability current* in Madelung's hydrodynamic interpretation [14], whereas the corresponding photon velocities along the trajectories are proportional to *weak (local) values of the canonical momentum* [15]. However, surprisingly, neither the original paper [1], nor numerous subsequent papers citing this work, mentioned the simple fact that Kocsis *et al*. measured nothing but the distribution of the *Poynting vector* in the optical interference field [16].

In this paper, we examine the experiment [1] from the viewpoint of classical optics. Kocsis *et al*. used the polarization of light as the measuring degree of freedom ("meter"), weakly coupled to the momentum of light via an anisotropic crystal. We show that, with the chosen polarization "pointer" (one of the Stokes parameters), Kocsis *et al*. measured the transverse Poynting-vector component in the electromagnetic interference field. At the same time, choosing another Stokes-parameter "pointer" would immediately result in the measurements of the *imaginary* part of the local photon momentum, sometimes called "osmotic velocity" [6,13,15,17]. Furthermore, we argue that the *same* measurements of the Poynting-vector distribution and weak values of the momentum are realized in classical optics via the motion of probe Rayleigh particles immersed in the field [18,19]. In this case, the particle represents a natural "meter" dipole-coupled to the field, and the particle's motion points along the local momentum of light. Although the coupling between one photon and the particle is *not weak*, the interaction with a multi-photon classical field yields the same weak value. We suggest that here it can be regarded as the expectation value of the photon momentum under the condition that it is found with a certain coordinate, based on the conditional probability [7]. Next, we discuss interesting situations where "anomalous" weak values of the momentum appear: either negative (so-called *backflows* [20]) or exceeding the wavenumber (superoscillations [4,21] and *superluminal* propagation [22,23]). We show that negative velocities accompanying optical vortices were described at least starting from the 1950s [24], whereas measurements of superluminal local velocities of photons have been realized in evanescent optical fields in the 1970s [25].

## 2. Classical description of the experiment by Kocsis *et al*.

The detailed quantum-mechanical analysis of the experiment by Kocsis *et al*. can be found in [1,15]. Therefore, we can now start with the classical-optics description of the measurements of the "average photon trajectories". As mentioned above, the single-photon or multi-photon character of the field does not make any difference after averaging over many events, and we describe the experiment as it would appear in the continuous-wave limit of classical optics.

The incident light in [1] represented a paraxial, monochromatic, uniformly-polarized wave propagating along the $z$-axis and modulated along the transverse *x*-coordinate. The complex electric field of such wave in the two-dimensional $\mathbf{r} = (x,z)$ geometry can be written as

$$\mathbf{E}(\mathbf{r},t) = \mathbf{e}\psi(\mathbf{r})\exp(-i\omega t), \tag{1}$$

where $\mathbf{e} = (e_x, e_y)$ is the transverse complex unit polarization vector ($\mathbf{e}^* \cdot \mathbf{e} = 1$, and we neglect the small longitudinal field component, $E_z \simeq 0$), whereas $\psi(\mathbf{r})$ is the complex scalar field, which slowly varies along the $x$-coordinate and behaves nearly as $\exp(ikz)$ along the $z$-coordinate. Here $k$ and $\omega = kc$ are the wavenumber and frequency, respectively; the common time dependence $\exp(-i\omega t)$ is omitted in what follows.

The wave (1) was transmitted through a thin plate of uniaxial calcite crystal, placed at some $z$-coordinate parallel to the $x$-axis. The optical axis of the crystal lies in the $(x,z)$ plane at an angle $\alpha_0$ from the $z$-axis. For an incident plane wave, calcite introduces a phase, which depends on the polarization $\mathbf{e}$ and direction of the wave vector $\mathbf{k}$ (thereby, providing a coupling between the polarization and momentum of the field [1]). The phase difference between the extraordinary ($x$-polarized) and ordinary ($y$-polarized) waves, $-\phi$, depends on the angle $\alpha$ between the $\mathbf{k}$-vector and the optical axis. In the paraxial approximation, $|k_x| \ll k$, one can write

$$\alpha \simeq \alpha_0 + \frac{k_x}{k}, \quad \phi(\alpha) \simeq \phi_0 + \zeta\frac{k_x}{k}, \tag{2}$$

where $\phi_0 = \phi(\alpha_0)$, $\zeta = \partial\phi(\alpha_0)/\partial\alpha_0$, and the parameters of the experiment were chosen such that $\phi_0 = 0 \bmod 2\pi$. An arbitrary paraxial wave (1) can be represented as a superposition of plane waves in the Fourier integral: $\mathbf{E}(x,z) = \int \tilde{\mathbf{E}}(k_x)\exp(ik_x x + ik_z z)dk_x$ (where $k_z = \sqrt{k^2 - k_x^2}$), and the $k_x$-dependent phase $-\phi$, Eq. (2), should be added to each Fourier component $\tilde{E}_x(k_x)$. This results in the perturbed transmitted field

$$E'_x(x,z) = \int \tilde{E}_x(k_x)\exp(ik_x x + ik_z z - i\phi)dk_x \simeq E_x(x - \Delta_x, z), \tag{3}$$

$$\Delta_x = \frac{\zeta}{k} = \frac{\partial\phi}{\partial k_x}. \tag{4}$$

Thus, the horizontal $x$-polarized component of the wave field experiences a lateral shift (4) with respect to the vertical $y$-polarized component. This is the usual birefringence of a uniaxial crystal. At the same time, Eq. (4) reveals its analogy to the Goos-Hänchen shift given by the Artmann formula for the $k_x$-dependent phase (2) of the transmission coefficient [26]. Interestingly, this analogy is not a coincidence: in the more general case of an $x$- and $y$-dependent field, the calcite plate with $\phi_0 \neq 0 \bmod 2\pi$ would produce the shifts $\Delta_x$ and $\Delta_y$, entirely analogous to the Goos-Hänchen and spin-Hall effect [9,10] shifts. In such three-dimensional case, the action of the crystal plate is described by a $k_x$- and $k_y$-dependent Jones matrix in the momentum (Fourier) representation [27].

Using the smallness of the shift (3) and (4) as compared with the large-scale $x$-modulation of the paraxial field, the resulting wave transmitted through the calcite plate can be written as

$$\mathbf{E}'(x,z) = \hat{\mathbf{x}}E_x(x - \Delta_x, z) + \hat{\mathbf{y}}E_y(x,z) \simeq \mathbf{E}(x,z) - \hat{\mathbf{x}}\Delta_x \frac{\partial E_x(x,z)}{\partial x}, \tag{5}$$

where $\hat{\mathbf{x}}$ and $\hat{\mathbf{y}}$ are the unit vectors of the corresponding axes. Equation (5) shows that the shift of the $x$-polarized field component also causes changes in the polarization, which now becomes slightly *non-uniform*. To characterize the polarization distribution in the input and transmitted fields, it is convenient to use the normalized Stokes-parameters vector $\vec{S} = (S_1, S_2, S_3)$ (it represents the polarization $\mathbf{e}$ on the Poincaré-Bloch sphere $|\vec{S}| = 1$):

$$S_1 = \frac{1}{I}\left(|E_x|^2 - |E_y|^2\right), \quad S_2 = \frac{1}{I}2\,\mathrm{Re}\left(E_x^* E_y\right), \quad S_3 = \frac{1}{I}2\,\mathrm{Im}\left(E_x^* E_y\right), \tag{6}$$

where $I(\mathbf{r}) = |\mathbf{E}(\mathbf{r})|^2 = |\psi(\mathbf{r})|^2$ is the field intensity distribution. In experiment [1], the input polarization was chosen as $\mathbf{e} = (1,1)/\sqrt{2}$, i.e., the Stokes vector was

$$\vec{S} = (0,1,0), \tag{7}$$

Calculating the Stokes vector (6) for the perturbed field (5), we arrive at

$$\vec{S}'(\mathbf{r}) \simeq \left(-\Delta_x \mathrm{Re}\frac{\partial \ln \psi(\mathbf{r})}{\partial x}, 1, \Delta_x \mathrm{Im}\frac{\partial \ln \psi(\mathbf{r})}{\partial x}\right). \tag{8}$$

Thus, the distributions of the two Stokes parameters, $S_3'(\mathbf{r})$ and $S_1'(\mathbf{r})$ (orthogonal to the input Stokes vector (7) on the Poincaré sphere) provide information about the real and imaginary parts of the $x$-component of the following complex vector:

$$\mathbf{p}(\mathbf{r}) = -i\nabla \ln \psi(\mathbf{r}) = \frac{\psi^*(\mathbf{r})(-i\nabla)\psi(\mathbf{r})}{|\psi(\mathbf{r})|^2}. \tag{9}$$

This "local momentum" resembles the expectation value of the canonical momentum operator $\hat{\mathbf{p}} = -i\nabla$, but without integration over space. In terms of weak measurements, Eq. (9) describes a complex "weak value of the photon momentum with the post-selection in the coordinate eigenstate" [1,15,16]:

$$\mathbf{p}(\mathbf{r}) \equiv \langle \mathbf{p}\rangle_{\mathrm{weak}} = \frac{\langle \mathbf{r}|\hat{\mathbf{p}}|\psi\rangle}{\langle \mathbf{r}|\psi\rangle}. \tag{10}$$

This can be regarded as the expectation value of the photon momentum under the condition that it is found at the point $\mathbf{r}$ [7]. For the paraxial fields under consideration, the $z$-component of the momentum (9) is known: $p_z(\mathbf{r}) \simeq k$. Taking this into account, the measurements of the Stokes-vector distributions (8) yield the complete information about the distribution of the local momentum (9) and (10) [28]. Note also that equation (8) for the Stokes parameters $\vec{S}'(\mathbf{r}) \simeq (\Delta_x \mathrm{Im}\, p_x(\mathbf{r}), 1, \Delta_x \mathrm{Re}\, p_x(\mathbf{r}))$ represents an example of the general weak-measurement situation where the "meter" is a qubit system [5,29].

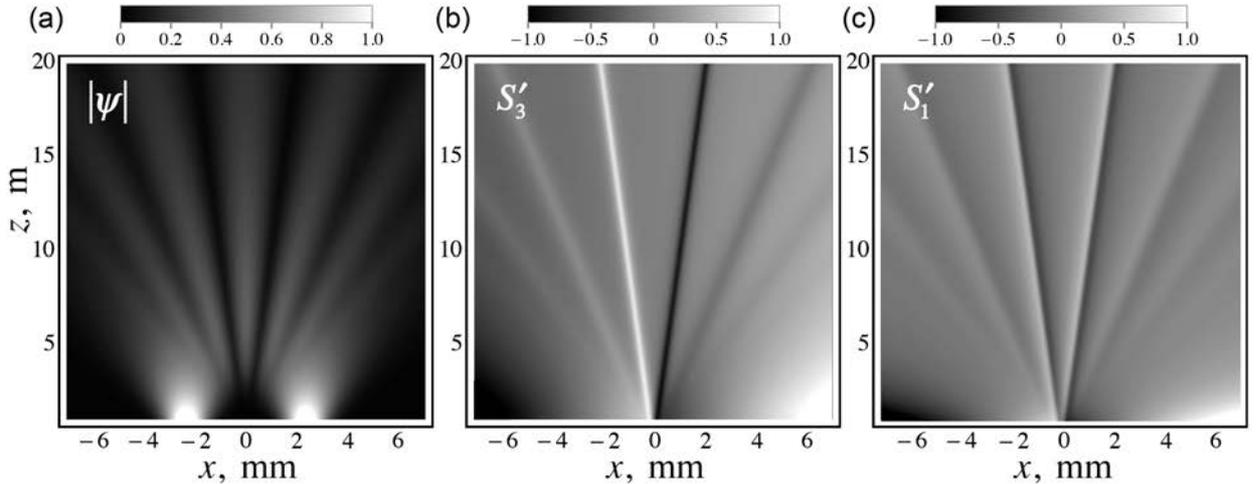

**Fig. 1.** Distributions of the wave amplitude $|\psi(\mathbf{r})|$ (a), and the Stokes parameters $S_3'(\mathbf{r}) \propto \mathrm{Re}\, p_x(\mathbf{r})$ (b) and $S_1'(\mathbf{r}) \propto \mathrm{Im}\, p_x(\mathbf{r})$ (c) for the interference of two Gaussian beams (12) with parameters [1] $\lambda = 2\pi/k = 0.943 \cdot 10^{-3}$ mm, $w_0 = 0.608$ mm ($z_R \simeq 1.23$ m), and $2a = 4.69$ mm. We depict a larger range of $z$ as compared with [1], in order to have a more pronounced picture of diffraction and currents.

Let us separate the real amplitude and phase of the wave function: $\psi(\mathbf{r}) = A(\mathbf{r})\exp[i\Phi(\mathbf{r})]$. Then, the *real* part of the complex local momentum (9) and (10) is equal to the gradient of the phase, i.e. the local wave vector [16]:

$$\text{Re}\,\mathbf{p}(\mathbf{r}) = \nabla\Phi(\mathbf{r}). \qquad (11a)$$

In the Madelung hydrodynamics [14] and Bohm mechanics [13], this quantity is proportional to the local velocity of the photons and, hence, the streamlines of the vector field $\text{Re}\,\mathbf{p}(\mathbf{r})$ determine the *photon trajectories*, which were depicted as the main result of [1]. In turn, the *imaginary* part of the local momentum, $\text{Im}\,\mathbf{p}(\mathbf{r})$, is equal to the gradient of the logarithm of the amplitude:

$$\text{Im}\,\mathbf{p}(\mathbf{r}) = -\nabla\ln A(\mathbf{r}). \qquad (11b)$$

This quantity is also extensively discussed in theory, and it is sometimes referred to as the "*osmotic velocity*" [6,13,15,17]. It is responsible for the appearance of the "quantum potential" in the Bohm theory [13]. Equation (8) shows that the experiment [1] could retrieve the distribution of $\text{Im}\,p_x(\mathbf{r})$ from the same weak measurements but with another "pointer": $S_1'(\mathbf{r})$ instead of $S_3'(\mathbf{r})$. The longitudinal component is negligible in paraxial fields: $\text{Im}\,p_z(\mathbf{r}) \simeq 0$.

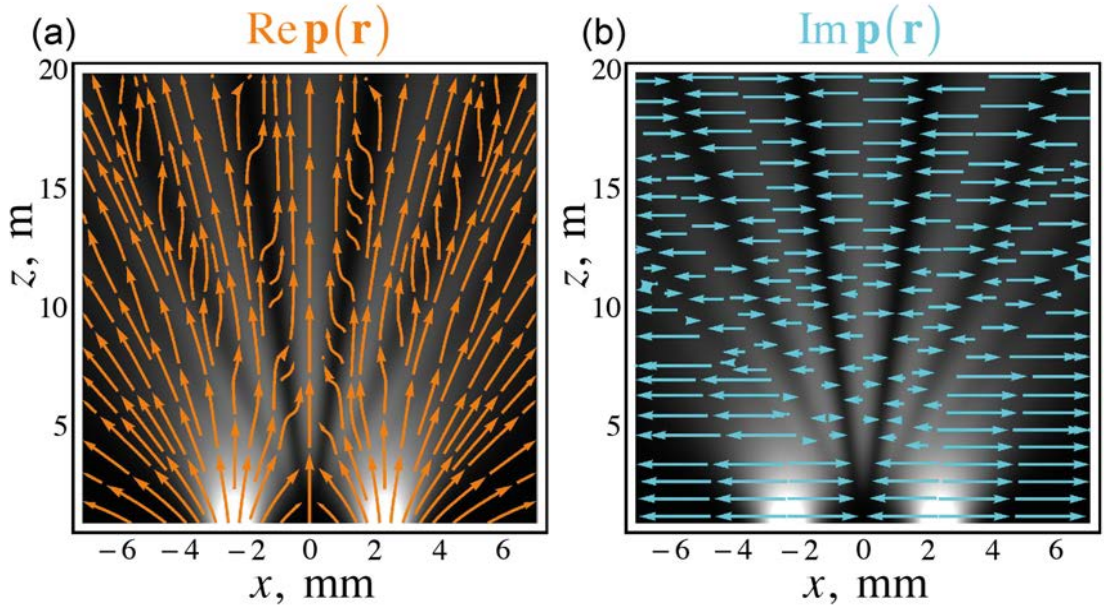

**Fig. 2.** Stream distributions of the real (a) and imaginary (b) parts of the local momentum $\mathbf{p}(\mathbf{r})$, Eqs. (9)−(11), for the two-beam interference of Fig. 1.

To illustrate distributions of the above quantities, we now consider the field used in experiment [1]. This is a superposition of two Gaussian beams propagating along the $z$-axis and mutually shifted along the transverse $x$-coordinate by a distance $2a$:

$$\psi = \frac{w_0}{w}\left\{\exp\left[-\left(\frac{1}{w^2} - \frac{ik}{2R}\right)(x-a)^2\right] + \exp\left[-\left(\frac{1}{w^2} - \frac{ik}{2R}\right)(x+a)^2\right]\right\}e^{ikz}. \qquad (12)$$

Here $w^2(z) = w_0^2(1 + z^2/z_R^2)$ is the beam width ($w_0$ being the waist), $R(z) = (z^2 + z_R^2)/z$ is the radius of curvature of the wavefronts, and $z_R = kw_0^2/2$ is the Rayleigh diffraction range. Figure 1(a) shows the amplitude distribution $A(\mathbf{r}) = |\psi(\mathbf{r})|$ of the field (12) (we plot the amplitude rather than the intensity for better contrast). Figures 1(b) and (c) depict the distributions of the measured Stokes parameters (8): $S_3'(\mathbf{r}) \propto \text{Re}\,p_x(\mathbf{r})$ and $S_1'(\mathbf{r}) \propto \text{Im}\,p_x(\mathbf{r})$. Supplementing these data with $p_z(\mathbf{r}) \simeq k$, in Figure 2 we plot the stream distributions of the corresponding real and imaginary parts of the local momentum (9)−(11): $\operatorname{Re}\mathbf{p}(\mathbf{r})$ and $\operatorname{Im}\mathbf{p}(\mathbf{r})$. Obviously, the streamlines of $\operatorname{Re}\mathbf{p}(\mathbf{r})$ in Fig. 2(a) yield the Madelung−Bohm trajectories known in the literature [13,16,30] and retrieved from the experimental measurements of $S_3'(\mathbf{r})$ in [1].

## 3. Poynting vector and weak measurements of momentum via probe particles

We are now in a position to show that the local momentum of photon, $\mathbf{p}(\mathbf{r})$, Eqs. (9) and (10), represents the *Poynting vector* normalized by the energy density (intensity). At first glance, this is not obvious as the time-averaged Poynting vector of a monochromatic electromagnetic field, $\mathbf{P}(\mathbf{r})$, has a rather different form:

$$\mathbf{P} = \frac{c}{2}\operatorname{Re}\left(\mathbf{E}^* \times \mathbf{B}\right). \tag{13}$$

Here $c$ is the speed of light in vacuum, $\mathbf{B}(\mathbf{r})$ is the complex wave magnetic field, and we use Gaussian units omitting inessential constants. However, it was recognized recently that the Poynting vector (13), representing the energy current, consists of two physically meaningful contributions: orbital (canonical) and spin currents [16,22,31,32]. Using the free-space Maxwell equations for a monochromatic field, Eq. (13) can be represented as

$$\mathbf{P} = \frac{c}{2\omega}\operatorname{Im}\left[\mathbf{E}^* \cdot (\nabla)\mathbf{E}\right] + \frac{c}{4\omega}\nabla \times \operatorname{Im}\left[\mathbf{E}^* \times \mathbf{E}\right] \equiv \mathbf{P}_\mathrm{O} + \mathbf{P}_\mathrm{S}, \tag{14}$$

where we use the notations of [16]: $\left[\mathbf{E}^* \cdot (\nabla)\mathbf{E}\right]_i = \sum_j E_j^* \nabla_i E_j$. Such separation of the canonical and spin parts of the energy or probability current is known in field theory and is common for any particles with spin [33]. Importantly, the spin current does not transport energy, because $\nabla \cdot \mathbf{P}_\mathrm{S} = 0$ and $\int \mathbf{P}_\mathrm{S}(\mathbf{r})dV = 0$, and it only generates the spin angular momentum of the field [31−33]. Therefore, it is the *orbital* part of the Poynting vector that should be associated with the canonical momentum density of the field. Moreover, for the linearly-polarized field used in [1], the spin contribution vanishes: $\mathbf{P}_\mathrm{S} = 0$, and the Poynting vector (13) identically coincides with its orbital part in Eq. (14): $\mathbf{P} = \mathbf{P}_\mathrm{O}$.

Taking the ratio of the canonical part of the Poynting vector, $\mathbf{P}_\mathrm{O}$, to the energy density $W = |\mathbf{E}|^2/2$, we obtain, for a uniformly-polarized field (1):

$$\frac{\mathbf{P}_\mathrm{O}(\mathbf{r})}{W(\mathbf{r})} = \frac{c^2 \operatorname{Im}\left[\psi^*(\nabla)\psi\right]}{\omega |\psi|^2} = \frac{c^2}{\omega}\operatorname{Re}\mathbf{p}(\mathbf{r}). \tag{15}$$

Thus, the ratio (15) represents the real part of the local momentum (9)−(11) with a constant pre-factor $c^2/\omega$. With such a pre-factor, it can be associated with the *local group velocity* of photons [22,23,34]. This can be seen in two ways: (i) in relativity, the particle velocity is given by $\mathbf{p}c^2/E$, where $\mathbf{p}$ and $E$ are the momentum and energy of the particle, and (ii) in the quantum-wave formalism, the group-velocity operator is $\hat{\mathbf{v}}_\mathrm{g} = \partial \omega(\hat{\mathbf{p}})/\partial \hat{\mathbf{p}} = c\hat{\mathbf{p}}/|\hat{\mathbf{p}}|$.

Thus, Eq. (15) shows that the experiment by Kocsis *et al.* [1] measured the transverse Poynting vector of the field. When normalized by the energy density (i.e., intensity, like it was also done for the Stokes parameters (6)), this quantity yields the weak value of the canonical momentum (10) or the local Madelung−Bohm velocity of photons [35]. At the same time, the Poynting vector is considered as an observable quantity in classical optics and it can be measured in a more straightforward (although, probably, less elegant) way. In hydrodynamics, a straightforward way to measure a current is to put a small *probe particle* into the flow and trace its motion. The same can be done in optics. Small particles experience the action of optical forces and move in light fields, which is widely used for optical manipulation [18].

Isotropic neutral particles, small compared with the wavelength, cause Rayleigh scattering of light. In this case, the particle-field interaction is typically approximated by the electric-dipole coupling [18,36]. The optical force acting on the Rayleigh particle consists of two contributions: a gradient force and a scattering force. For a particle with complex polarizability $\chi$, the optical force is given by [18,36]:

$$\mathbf{F} = \frac{1}{2}\operatorname{Re}\chi \operatorname{Re}\left[\mathbf{E}^* \cdot (\nabla)\mathbf{E}\right] + \frac{1}{2}\operatorname{Im}\chi \operatorname{Im}\left[\mathbf{E}^* \cdot (\nabla)\mathbf{E}\right] \equiv \mathbf{F}_{\text{grad}} + \mathbf{F}_{\text{scat}}. \tag{16}$$

Remarkably, the two forces in Eq. (16), normalized by the energy density $W = |\mathbf{E}|^2/2$, are proportional to the imaginary and real parts of the local momentum (9)−(11). Hence, observing the motion of the particle with real and imaginary polarizabilities, one can measure the complex local momentum $\mathbf{p}(\mathbf{r})$. In particular, the scattering force 'measures' the canonical part of the Poynting vector (14) and the Madelung−Bohm velocity of photons (11a), while the gradient force 'measures' the non-normalized imaginary part of the local momentum, i.e., the "osmotic velocity" of the photons (11b).

Thus, the force acting on a particle (normalized by the wave intensity) yields the same weak value (10) of the photon momentum [16]. This demonstrates the universality of weak values, which are independent of the choice of the "meter", i.e., the measuring system. In the experiment [1], the polarization played the role of the "meter" (the Stokes parameters being the "pointers"), whereas the anisotropic calcite crystal provided a weak coupling between the polarization and momentum of light. In the case of probe particles, the particle is the "meter" (with the forces or their manifestations being the "pointers"), and the coupling is provided by the electric-dipole interaction between light and matter. Note that this coupling is not weak (as the photons are absorbed or strongly scattered by the particle), and this classical experiment cannot be immediately interpreted in terms of quantum weak measurements [3−5], i.e., with weak measurement of the momentum and subsequent strong measurements (post-selection) of the coordinate. Nonetheless, the perturbation of the classical multi-photon field is weak, and there is some similarity with quantum weak measurements. On the one hand, a heavy and well-localized particle strongly measures coordinates even for a single absorbed photon, i.e., provides post-selection. On the other hand, it weakly measures momentum: the impact of a single photon on the particle's motion is negligible, but the averaged impact of multiple photons is noticeable. Therefore, such measurement represents the average (expectation) value (10) of the photon momentum given that the photon is found at the point with coordinate $\mathbf{r}$. Apparently, this can be described using the corresponding conditional probability [7] and without involving the quantum weak-measurement scheme [37].

Experimentally, tracing the motion of probe particles to detect streamlines of optical currents is a challenging problem, as compared to the elegant method of [1]. Nonetheless, this technique was successfully used a decade ago for the detection of the swirling Poynting vector in optical *vortex beams* [19] and in various complicated optical fields [38]. Let us consider the simplest example of a vortex beam – a Bessel beam propagating along the $z$-axis [39]. For paraxial uniformly-polarized beams, $\mathbf{E}(\mathbf{r}) = \mathbf{e}\psi(\mathbf{r})$, the problem is insensitive to the polarization $\mathbf{e}$, and we consider the scalar field function:

$$\psi(r,\varphi,z) = J_{|\ell|}(k_\perp r)\exp(i\ell\varphi + ik_z z). \tag{17}$$

Here $(r,\varphi,z)$ are cylindrical coordinates, $J_{|\ell|}$ is the Bessel function of the first kind, $\ell$ is the integer vortex charge, and $k_\perp \ll k$ and $k_z = \sqrt{k^2 - k_\perp^2}$ are the transverse and longitudinal wavenumbers. Substituting Eq. (17) into Eqs. (14) and (9)−(11), we obtain the orbital part of the Poynting vector and normalized local momentum of the field:

$$\mathbf{P}_\text{O} = \frac{c^2}{\omega}|J_{|\ell|}(k_\perp r)|^2 \left(\frac{\ell\hat{\boldsymbol{\varphi}}}{r} + \hat{\mathbf{z}}k_z\right), \quad \mathbf{p}(\mathbf{r}) = \left(\frac{\ell\hat{\boldsymbol{\varphi}}}{r} + \hat{\mathbf{z}}k_z\right) - i\hat{\mathbf{r}}\frac{\partial \ln|J_{|\ell|}(k_\perp r)|}{\partial r}, \tag{18}$$

where $\hat{\mathbf{r}}$, $\hat{\boldsymbol{\varphi}}$, and $\hat{\mathbf{z}}$ are the unit vectors of the cylindrical coordinates. Equation (18) shows that the Poynting vector and local momentum spiral in Bessel beams, and the "averaged photon trajectories" (streamlines of these currents) are *helixes* [40].

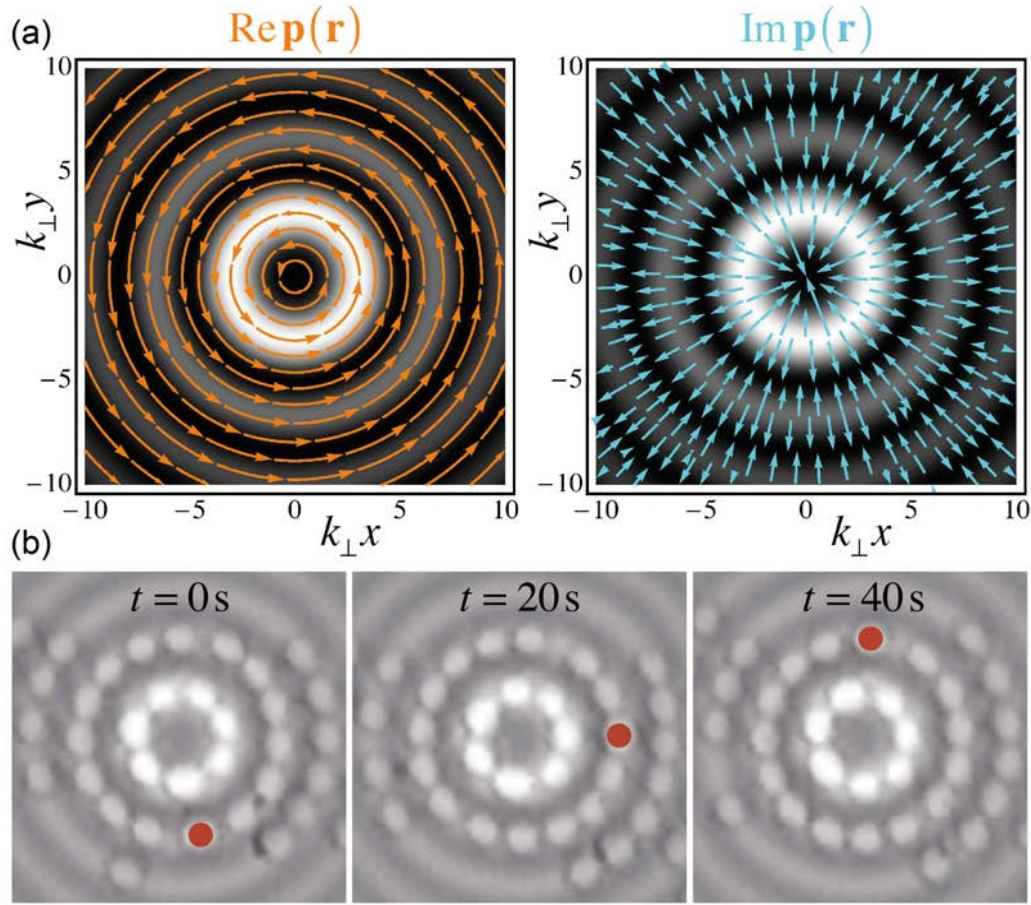

**Fig. 3.** (a) Intensity (grayscale background) and stream distributions of the real and imaginary parts of the normalized local momentum $\mathbf{p}(\mathbf{r})$, Eqs. (9)−(11), in the transverse plane of the Bessel beam (17) with $\ell = 2$. (b) Successive frames image the motion of the probe particles in the Bessel-beam field (taken from [19a] and one particle is marked by the red circle). A movie of this motion, available in [19a], shows that the inner rings move faster because the radiation force is proportional to the field intensity and to $1/r$. The particles are trapped at the radial field maxima because of the gradient force anti-parallel to the "osmotic velocity" $\text{Im}\,\mathbf{p}(\mathbf{r})$.

In Figure 3 we show the stream distributions of the real and imaginary parts of the local momentum (18) in the transverse $(x,y)$ plane of the Bessel beam with $\ell = 2$. This is compared with experimental images taken from Ref. [19a] and showing circular motion of $3\mu$m silica spheres along the streamlines of the Poynting vector [due to the scattering force in Eq. (16)], as well as radial attraction of particles to the intensity maxima due to the gradient force in Eq. (16). Since the amplitude of the transverse Poynting vector and radiation force is proportional to the field intensity and to $1/r$, the rotational velocities of the particles decrease for higher radial maxima of the Bessel beam; this can be seen in the online movie supplementing [19a]. Normalizing the particle velocities by the local field intensity yields the real part of the transverse local momentum, $\text{Re}\,\mathbf{p}_\perp(\mathbf{r}) = \ell\hat{\boldsymbol{\varphi}}/r$. Akin to [1], adding the known longitudinal component $p_z = k_z$, one can reconstruct the three-dimensional helical trajectories of photons, i.e., streamlines of $\text{Re}\,\mathbf{p}(\mathbf{r})$. They are given by the equations $\varphi(z) = \varphi_0 + z\ell/k_z r_0^2$ and $r(z) = r_0$ [40c], and are shown in Fig. 4. Note that relatively large micron-size particles were used in [19a], while modern optical experiments allow using metallic nano-particles. This would provide a much higher resolution for measurements of the Poynting currents. Furthermore, other types of optical "meters" sensitive to the direction of the energy current can also be used. In particular, local measurements of the Poynting vector have been realized using Shack–Hartman sensors [41], moiré deflectometry [42], and near-field fiber probes [43]. These techniques should also be analyzed from the viewpoint of the momentum weak values.

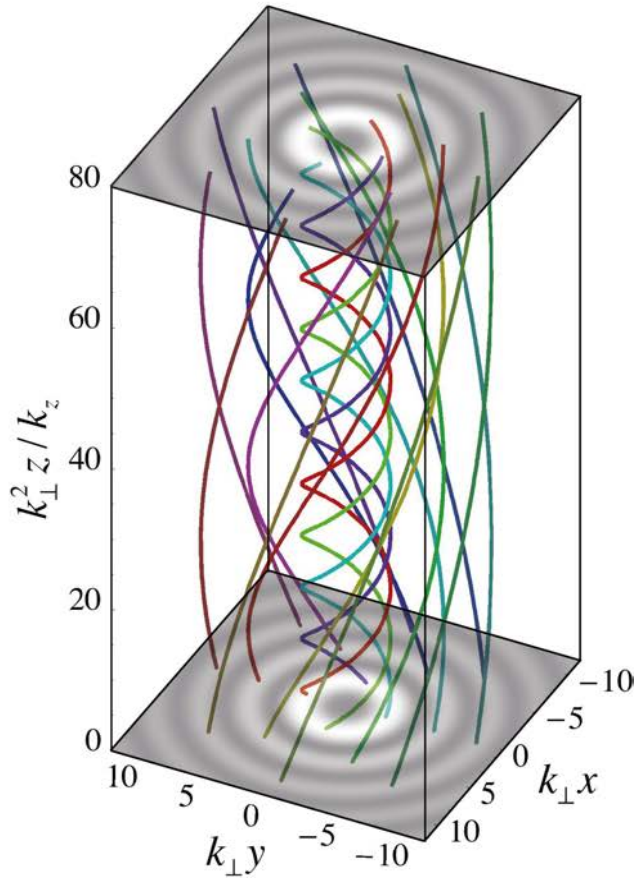

**Fig. 4.** Three-dimensional "trajectories of photons", i.e., streamlines of $\text{Re}\,\mathbf{p}(\mathbf{r})$ (18) for the Bessel beam (17) with $\ell = 2$. Akin to [1], these trajectories can be reconstructed from measurements of the transverse Poynting-vector distribution, Fig. 3, supplemented by the constant longitudinal component $p_z = k_z$.

## 4. Anomalous local velocities: backflows and superluminal propagation

It was noticed by Berry [16] that a real part of the local momentum $\text{Re}\,\mathbf{p}(\mathbf{r})$, Eqs. (9)−(11), can take "anomalous" values as compared to the momentum spectrum (wave vectors) of the field. In particular, $\text{Re}\,\mathbf{p}(\mathbf{r})$ diverges in the vicinity of optical vortices (phase singularities), where $|\psi(\mathbf{r})| \to 0$. This is seen in the second Eq. (18) with $|\text{Re}\,\mathbf{p}_\perp(\mathbf{r})| = 1/r \to \infty$ at $r \to 0$. This is the phenomenon of "superoscillations" [4,21], where the function changes faster than any component of its Fourier spectrum: $|\text{Re}\,\mathbf{p}(\mathbf{r})| > k$ is our case. Superoscillations represent an example of quantum weak values [3−5]: the weak value of the momentum (10) lies outside the momentum spectrum.

Since we consider monochromatic waves propagating in the positive $z$-direction, the spectrum of $\hat{p}_z$ is $k_z \in [0,k]$, and one can find two types of "anomalous" values of $\text{Re}\, p_z(\mathbf{r})$: either negative or exceeding $k$. Firstly, *negative* local momentum, $\text{Re}\, p_z(\mathbf{r}) < 0$, represents optical or quantum "*backflow*" [20]. Berry showed [20c] that regions of optical backflows are attached to *vortices*; this is natural as the vicinity of the vortex up to the stagnation point contains all directions of the current. Various examples of vortex backflows in the basic scattering, focusing, and diffraction problems are described in optics at least starting from the 1950s [24]. For instance, Wolter [24a] considered streamlines of the Poynting vector in the total internal reflection of a wave packet (the Goos-Hänchen problem). He showed that although all plane waves in the field spectrum propagate in the same $z$-direction along the glass-air interface $x = 0$, the Poynting vector exhibits a vortex in the glass, which inevitably has an area with backflow $\text{Re}\, p_z(\mathbf{r}) < 0$ [44] (see Fig. 5 below). Note that the Wolter's paper was titled "Concerning the path of light upon total reflection" (cf. "average trajectories of photons" in [1]). Wolter modeled "wave packet" by only two plane waves in the spectrum. The same vortex in the total internal reflection of a proper wave packet, including the imaginary current $\text{Im}\,\mathbf{p}(\mathbf{r})$, was described in a profound paper by Hirschfelder *et al.* [17b]. In modern optics, the retrograde Poynting vectors are discussed for potential application in the so-called "tractor beams", which could transport particles in the direction opposite to the beam propagation [45]. However, the typical size of the backflow areas is smaller than the wavelength, and this could barely lead to an efficient transport.

Secondly, an anomalously *large* local momentum occurs if $\text{Re}\, p_z(\mathbf{r}) > k$. This again appears due to *superoscillations* in the vicinity of vortices where $|\text{Re}\,\mathbf{p}(\mathbf{r})| \to \infty$ [16,21], but also in *evanescent* waves [22]. According to (15), the quantity

$$\frac{c}{k}\text{Re}\,\mathbf{p}(\mathbf{r}) = \mathbf{v}_g(\mathbf{r}) \equiv \langle \mathbf{v}_g \rangle_{\text{weak}} \qquad (19)$$

represents the weak (local) value of the *group velocity* of light [22,23], which coincides with the Madelung−Bohm velocity of photons. Then, the anomalously large local momentum $\text{Re}\, p_z(\mathbf{r}) > k$ corresponds to the *superluminal* local group velocities $v_{gz} > c$. Superluminal velocities were recently described in superoscillatory areas of random propagating waves [21e,23] and in evanescent waves [22].

Let us consider a single evanescent wave propagating along the $z$-axis and decaying along the $x$-axis:

$$\psi(\mathbf{r}) = \exp(ik_z z - \kappa x), \quad v_{gz} = ck_z/k > c. \qquad (20)$$

The latter inequality follows from the dispersion relation $k_z^2 - \kappa^2 = k^2$ and shows that the local group velocity (19) is *always superluminal* in evanescent waves [22,46]. However, a single evanescent wave (20) cannot exist in unbounded free space. Consider, for example, a simple localized solution involving evanescent waves: a surface plasmon-polariton propagating along an air-metal interface $x = 0$ [47]. In the $x > 0$ (air) half-space, the scalar wave function is given by (20), whereas in the $x < 0$ (metal) half-space, it is another evanescent wave decaying in the negative $x$-direction. Importantly, the ratio of the Poynting vector integrated over the whole $x$-axis, to the integral energy yields the *strong value* of the group velocity, which is always *subluminal* [46,48]:

$$\frac{\int P_{Oz}(x)dx}{\int W(x)dx} = \langle v_{gz} \rangle_{\text{strong}} < c. \qquad (21)$$

Thus, measuring the local momentum $p_z = k_z > k$ in an evanescent wave, one performs a measurement of the weak value of momentum (10) or superluminal group velocity (19) and (20), with the post-selection of photons in a classically-forbidden zone. As described in Section 3, such measurements can be realized using radiation force (16) on a small probe particle.

Remarkably, somewhat similar measurements were made in 1978 by Huard and Imbert [25] using an interaction and momentum exchange between a moving atom and evanescent wave from the total internal reflection. The Doppler-shift measurements of [25] detected the momentum transfer from the evanescent wave to the atom larger than $k$. Although the measurement of the photon momentum via resonant interaction and Doppler shift is strong and causes delocalization of the atom, it is assured that the atom is located in the evanescent-wave half-space $x > 0$, which is sufficient for post-selection and measurements of the weak value (19) and (20). Hence, the experiment by Huard and Imbert could be regarded as the first measurement of the anomalous weak values of the photon momentum and superluminal local group velocity.

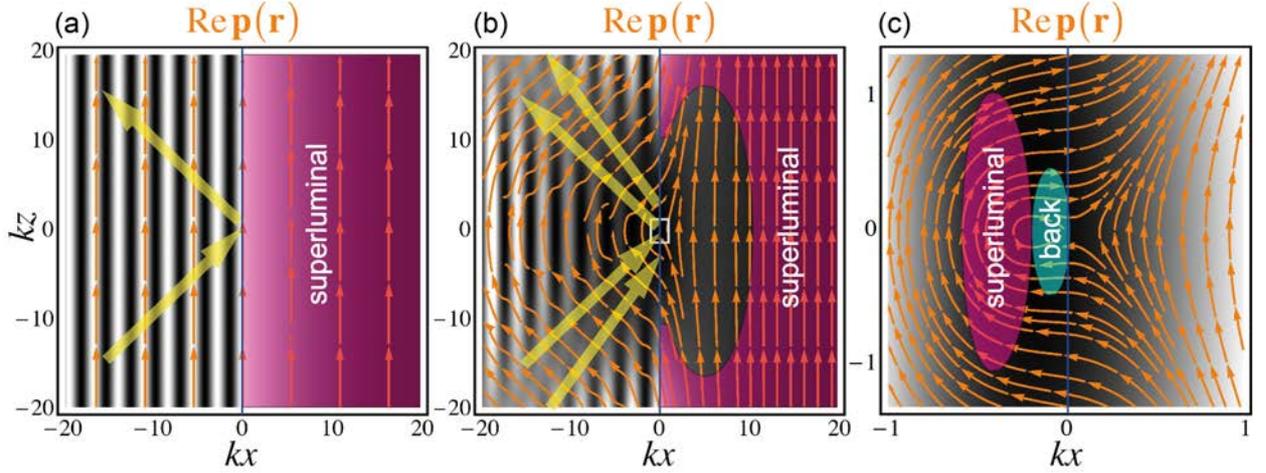

**Fig. 5.** Intensities (grayscale background) and stream plots of the local momentum $\text{Re}\,\mathbf{p}(\mathbf{r})$ for light undergoing total internal reflection at the interface $x = 0$ between glass ($x < 0$) and air ($x > 0$). The areas of the superluminal and backward momenta $\text{Re}\,p_z(\mathbf{r})$ are indicated by purple and cyan, respectively. We consider transverse $y$-polarization and parameters used in [24a,44]. (a) A single plane wave is incident and reflected as shown by yellow arrows. This produces an evanescent wave (20) in the air, with superluminal momentum $\text{Re}\,p_z > k$, which was measured by Huard and Imbert using atomic Doppler shift [25]. (b) Two plane waves are incident and reflected as considered by Wolter [24a]. The emerging vortex slows down the superluminal current in the evanescent wave. (c) Zoom-in view of the subwavelength vortex vicinity [indicated by the tiny rectangle in (b)] shows that areas of the superluminal ($\text{Re}\,p_z(\mathbf{r}) > nk$) and backward ($\text{Re}\,p_z(\mathbf{r}) < 0$) currents appear in the glass.

To illustrate the anomalous values of the local momentum, $\text{Re}\,p_z(\mathbf{r}) \notin (0, k)$, and their appearance near vortices and evanescent waves, we consider the total internal reflection of light at a glass-air interface $x = 0$, Fig. 5. When a single plane wave is incident from the glass ($x < 0$) with $k_z > 0$, it generates a uniform evanescent wave in the air ($x > 0$), Eq. (20). Although the propagating waves in the glass and evanescent wave in the air have the same local momentum $\text{Re}\,p_z = k_z$, it is subluminal in the glass with refractive index $n$ ($\text{Re}\,p_z < nk$) and superluminal in the air ($\text{Re}\,p_z > k$), see Fig. 5(a). In order to construct a wave packet or beam, one has to consider multiple plane waves incident at slightly different angles. The simplest model of two incident plane waves was considered by Wolter [24a,44], and is shown in Figs. 5(b) and (c). One can see that the destructive interference of the two incident and two reflected waves produces a vortex in the glass, near the interface. A magnified view of the subwavelength details of the vortex, Fig. 5(c), shows that the vicinity of the vortex contains areas of superluminal ($\text{Re}\,p_z(\mathbf{r}) > nk$) and backward ($\text{Re}\,p_z(\mathbf{r}) < 0$) local momenta.

# 5. Discussion

We have examined the weak-measurements approach to the detection of the local momentum (velocity) and corresponding Madelung−Bohm trajectories of photons. Since weak measurements imply averaging over many events, the measured quantities represent well-known characteristics of classical optical fields. Namely, the real part of the weak value of photon momentum represents the orbital part of the Poynting vector, normalized by the energy density. In this manner, the Madelung−Bohm trajectories of photons naturally coincide with the streamlines of the Poynting vector.

We revisited recent experiment by Kocsis *et al*. [1] and provided its classical-optics description. The same experiment could measure not only the real but also the imaginary part of the local momentum of light, the so-called "osmotic velocity". Next, we showed that the forces acting on probe Rayleigh particles in optical fields allow measurements of the real and imaginary parts of the same local momentum, in a way similar to investigations of hydrodynamical flows. As an illustration, we considered the Poynting vector and trajectories in Bessel beams, observed in optical experiments [19] using probe particles.

Weak (local) values of the momentum can be "anomalous", i.e., can lie outside of the Fourier spectrum of the field. For a monochromatic field propagating in the positive $z$-direction, the spectrum of $\hat{p}_z$ lies in the $[0,k]$ range. Accordingly, the anomalous weak values can be either negative (optical backflow) or exceeding the wavenumber (superoscillations and superluminal propagation). We demonstrated the relation of the anomalous local velocities to both optical vortices and evanescent waves. Remarkably, the optical-vortex backflows are described in basic optical problems starting from the 1950s [24], while the superluminal local momentum in evanescent waves was experimentally measured in 1978 by Huard and Imbert [25]. The latter can be regarded as a measurement of the anomalous weak value of the photon momentum (or group velocity) with the post-selection in a classically-forbidden zone.

It should be emphasized that we do *not* oppose the classical description to the quantum weak-measurement formalism. On the contrary, we hope that the classical picture complements it, giving additional insight and revealing interesting links. Importantly, the weak-measurement approach appears as a great unifying concept. Only within this paper and particular problem, it readily brings together the Madelung−Bohm trajectories in the two-slit interference, the Goos-Hänchen effect, optical vortices, interaction with particles, and evanescent waves. Note also that although the experiment [1] used rather standard and simple optical tools, without weak measurements one never considered a uniaxial crystal and polarizers as an instrument for the measurements of the local Poynting vector.

Let us briefly mention some problems which are left outside of the present consideration but could be important in a future analysis. Firstly, backflows of the probability current can be associated with *negative probabilities* [20], which appear in the theory of weak measurements [3−5,7]. (In the case of light, where the energy rather than the probability current occurs, one should regard negative frequencies.) Therefore, it is tempting to examine a possible connection of retrograde currents with negative-probability effects, such as the "quantum three-box paradox" [49] or the weak-valued momentum transfer probabilities measured in a twin-slit experiment [50].

Secondly, the appearance of vortices in many weak-measurement problems hints at the possible fundamental role of vortex structures [51]. Both superoscillatory and backflow regions appear near vortices [16,20c,21,23]. At the same time, vortices are related to the *angular momentum* of the field, which make the wave essentially *non-local* (an object carrying angular momentum cannot be shrunk into a point [52]). Possible connections between vortices, angular momentum, backflows, on the one side, and non-locality, causality in quantum processes, on the other side, represent an intriguing problem for investigations.

Finally, in this paper, for the sake of simplicity, we considered only the *electric* wave field and its properties. However, free-space fields inevitably include *electric* and *magnetic* fields in a symmetric form. This is an important "dual symmetry" responsible for the conservation of the helicity of photons [16,32,53]. Accordingly, all fundamental conserved quantities, such as canonical and spin currents in Eq. (14), should contain an arithmetic average of the electric and magnetic parts [16,32]. These parts are equivalent in paraxial fields, but different in non-paraxial waves, such as Bessel beams and evanescent waves [16,22]. At the same time, experimental measurements of the proper dual-symmetric quantities represent a challenging problem. The point is that any optical measurements involve *light-matter interactions*, where matter acts as a meter. However, properties of matter are fundamentally electric-magnetic *asymmetric*: there are electric but not magnetic charges. Therefore, most of detectors (including uniaxial crystals and Rayleigh particles, considered here) are sensitive only to the electric parts of the proper electric-magnetic properties of the fields. Here we face a general problem that any quantum measurement represents an interaction, which can crucially depend on the properties of the meter and character of interaction. From the field-theory point of view both light and matter represent two parts of one system, which are conditionally separated in the quantum-measurement approach. We discuss these issues in more details elsewhere [32].

We are grateful to M.V. Berry, M.R. Dennis, K. Dholakia, A. Dogariu, H.F. Hoffman, and S.K. Ozdemir for fruitful and stimulating discussions. This work was partially supported by the European Commission (Marie Curie Action), ARO, JSPS-RFBR contract No. 12-02-92100, Grant-in-Aid for Scientific Research (S), MEXT Kakenhi on Quantum Cybernetics, and the JSPS via its FIRST program.